# Fermiology and Band Structure of Oxygen-Terminated $Ti_3C_2T_x$ MXene

Martin Magnuson[1,*], Per Eklund[1,2,*], and Craig Polley[3]

[1]*Department of Physics, Chemistry and Biology (IFM), Linköping University, SE-581 83 Linköping, Sweden*
[2]*Inorganic Chemistry, Department of Chemistry - Ångström, Uppsala University, Box 538, SE-751 21 Uppsala, Sweden*
[3]*MAX IV Laboratory, Lund University, Fotongatan 2, SE-22484 Lund, Sweden.*

*martin.magnuson@liu.se ; per.eklund@kemi.uu.se

The class of two-dimensional carbides and nitrides known as MXenes exhibit remarkable electronic properties. Tailoring these properties, however, requires in-depth understanding of the band structure and Fermi-surface topology. Surface oxidation of MXenes has previously hampered the characterization of their Fermi surface, which is crucial for understanding the topology and anisotropy in the electronic structure and ultimately for tailoring electronic properties. Here, we reveal the Fermi surface topology and band structure of purely oxygen-terminated $Ti_3C_2T_x$ MXene achieved through rigorous thin film sample preparation and ultrahigh vacuum annealing. Polarized synchrotron radiation-based angle-resolved photoemission spectroscopy reveals electron pockets, bulk bandgaps, and a Dirac-like feature in the anisotropic electronic band structure. This paves the way for a fundamental understanding of band engineering of electronic transport properties, providing insights of importance for energy storage devices, transparent conductors, and catalysis.

The quantum phenomena occurring in graphene [1] has sparked significant interest in the exploration of two-dimensional (2D) materials with diverse electronic and optical properties that do not exist in bulk materials. 2D crystals with trigonal symmetry, *e.g.*, with hexagonal honeycomb structure, are known to host Dirac fermions at the corners of the hexagonal Brillouin zone (BZ), denoted Dirac points, conical band crossings corresponding to a zero bandgap [2]. There, charge-carriers behave as massless fermions, which has led to the emergence of Dirac physics, inducing new physical phenomena such as Klein tunneling and Zitterbewegung, and of utmost importance for topological insulators [3,4]. Beyond elemental 2D materials, such as graphene, silicene [5], most recently goldene [6], and binary compounds like hexagonal boron nitride (h-BN), molybdenum disulfide ($MoS_2$), tungsten disulfide ($WS_2$), there are great prospects for the large class of tunable 2D materials known as MXenes ($M_{n+1}X_nT_x$), where M represents a transition metal, X can be either carbon or nitrogen, and T denotes termination species [7]. These materials offer a wide range of compositional variables that can be manipulated to achieve specific properties [8], for possible applications as transistors and spintronics [9], 2D-based electronics and displays [10], supercapacitors [11], and transparent conductive electrodes [12].

In $M_{n+1}X_nT_x$, where n = 1, 2, or 3, the n+1 M and n X monolayers are arranged in an alternating sequence, as depicted in Figure 1(a). The properties are largely determined by the terminating species, including fluorine, hydroxide, chlorine, and atomic oxygen, for applications in semiconducting materials and electrochemical energy storage [13]. To prevent property deterioration [14], caution must be exercised in avoiding oxidation of the MXene surfaces. From density functional theory (DFT), there are many studies of band structures in MXenes and how they depend on termination, primarily relying on ground state DFT calculations at 0 K [15-17]. These calculations indicate that bare $M_{n+1}X_n$ layers exhibit metallic characteristics, with a finite density of states at the Fermi level ($E_F$), while $M_{n+1}X_nT_x$ can display semiconducting behavior [15-17] that can be altered by different termination species. However, consensus regarding the specific species inducing semiconducting behavior remains elusive. Of particular interest is the prediction of Fashandi *et al.* [18] who predicted that under certain circumstances, $Ti_3C_2T_x$ could exhibit Dirac points with giant spin-orbit splitting around the $E_F$, which is further supported by simliar predictions by others [19-21].

Despite this interest in MXenes, there has been limited effort directed towards comprehending the fundamental properties of their Fermi surfaces (FS), known as fermiology, and its effect on the material characteristics. Angle-resolved photoemission spectroscopy (ARPES) is a powerful experimental method to investigate FSs [22] but the surface must be perfectly clean and well ordered. This condition was not previously met for MXenes, as surfaces are often disordered and contaminated with a mixture of organics and residues from the synthesis. Schultz *et al.* [23] applied ARPES on delaminated $Ti_3C_2T_x$, but the terminations were mixed, and the data was azimuthally averaged due to randomly oriented flakes in the measured film, hindering the extraction of the shape and size of the FS.

Here, we investigate the electronic band structure and FS of a thin film stacked-nanosheet $Ti_3C_2O_x$ sample with pure oxygen termination obtained from etched $Ti_3AlC_2$-phase deposited on a *c*-axis-oriented sapphire ($Al_2O_3$) substrate (see Supplementary Information). By utilizing synchrotron-radiation-based ARPES at various excitation energies, we reveal inherent 2D features within the electronic structure and map the FS topology, including the orbital characteristics of the band structure and probing the broad momentum and energy range across the entire Brillouin zone. We provide direct experimental evidence for conical band crossings, analogous to Dirac points, at 1.5 eV binding energy below the $E_F$.

Sample preparation was performed to achieve oxygen as sole termination species [24] (see SI). Fig. 1(b) shows the evolution of low electron energy diffraction (LEED) spots during heat cycles in the form of high temperature flashes described in the SI. XPS surface characterization showed that the surface is exclusively oxygen-terminated (Fig. S5) and the final sharp LEED spots indicate a clean well-ordered hexagonal MXene surface. Fig. 1(c) shows four adjacent Brillouin zones (BZ) with high symmetry points at: $(\frac{2}{3}, \frac{1}{3}, 0)$ indicated along the two dashed high-symmetry lines; Fig. 1(d) shows models of a 1x1 unit cell and a 2x2 structure. The 2D hexagonal surface Brillouin zone (SBZ) of $Ti_3C_2O_x$ has the distances between symmetry points: $\bar{\Gamma}$-$\bar{M}$ = $2\pi/a\sqrt{3}$ = 1.180 Å$^{-1}$, $\bar{\Gamma}$-$\bar{K}$ = $\bar{\Gamma}$-$\bar{M}/\cos(30°)$ = 1.362 Å$^{-1}$, $\bar{K}$-$\bar{M}$ = $\tan(30°)\bar{\Gamma}$-$\bar{M}$ = 0.681 Å$^{-1}$.

Figure 2 shows ARPES measurements at 180 eV photon energy of $Ti_3C_2O_x$ along the $\bar{M}$-$\bar{\Gamma}$-$\bar{M}$ (top panel) and the $\bar{K}$-$\bar{\Gamma}$-$\bar{K}$ (bottom panel) direction. The shape of the overlaid calculated bands for the $Ti_3C_2O_2$ unit cell are generally in good agreement with the measured overview ARPES data. Near $E_F$, the valence band mainly consists of metallic Ti *3d*-$t_{2g}$ states, that provide the dominant contribution to conduction. Between 0.5-2 eV binding energy, a bulk bandgap is narrowest in the vicinity of the $\bar{\Gamma}$-point where there are highly dispersing bands. The absence of any $k_z$ dispersion when varying the photon energy are a clear indication of the 2D character of the metallic bands at $E_F$ (see SI). Below the bandgap, there are bonding states between Ti *3d*-$e_g$ and C *2p* orbitals [17,21]. At higher binding energy ($E_b$), there is a belt of flat bands with high intensity down to 8 eV, where a second bulk bandgap occurs in the region between 9-10.5 eV.

Figure 3(a) shows a magnification of the dispersion map around the $\bar{\Gamma}$-point where the horizontal dashed lines, denoted (b), (c), (d) at binding energies 0 eV, 1 eV, and 2 eV below the $E_F$, are constant energy contour cuts from the measured volume data. The FS of $Ti_3C_2O_x$ in Fig. 3(b) exhibits an intense pocket of highly dispersing light electron bands centered at the $\bar{\Gamma}$ point with heavier bands forming radial pockets pointing towards the $\bar{K}$-points at the corners of the BZ. Between the radial electron pockets, there are lens-shaped lobes of dark empty areas pointing towards the $\bar{M}$-points. Hence, the FS has a hexagonal warping with six-fold symmetry, centered around the $\bar{\Gamma}$-point of the BZ. The topology of the FS is thus very sensitive to the exact position of the $E_F$, that depends on the coverage and type of $T_x$. At the $E_F$, hole-like Ti $3d_{xy}$, heavy hole bands govern the conductivity along the basal plane, while electron-like Ti $3d_{z2}$ bands at the $\bar{\Gamma}$-point govern the conductivity along the *c*-axis (S1).

In contrast to the FS, the surface map at 1 eV binding energy below $E_F$ (Fig. 3c), displays symmetrical six-fold elliptical lobes of heavy electron bands extending radially from the center of the BZ towards the $\bar{M}$-points with a dark area around the $\bar{\Gamma}$-point that signifies a band gap. At 2 eV below the $E_F$, the energy cut in Fig. 3(d) exhibits weaker electron pockets pointing towards the $\bar{M}$-points.

Figure 4 shows polarization-dependent dispersion maps at a photon energy of 40 eV along the $\bar{K}$-$\bar{M}$-$\bar{K}$ BZ edge with (a) horizontal and (b) vertical polarization. With horizontal polarization (a), O $2p_z$ and Ti $3d_{z^2}$ orbitals are mainly probed perpendicular to the basal *ab*-plane. The two intense spots at 5.4 eV binding energy halfway along the $\bar{K}$-$\bar{M}$ direction are associated with both $O_{fcc}$ (oxygen occupying a surface fcc-site) and $O_b$ $2p_z$ orbitals [24]. In case (b), an elongated intense spot is centered at the $\bar{M}$-point at 5.2 eV. Here, the highly dispersing bands exhibit a shape of an hourglass, like a Dirac point, with a band crossing at 1.5 eV The highest intensity around the $\bar{M}$-point is attributed to O $2p_{xy}$ orbitals in the basal *ab*-plane. With the $\bar{E}$-vector aligned perpendicularly to the $Ti_3C_2O_x$ surface, there would only be intensity from O in the points halfway between the $\bar{M}$ and the $\bar{K}$ and thus only an $O_{fcc}$ contribution probed. Thus, for the vertical *p*-polarization (b), when the photon $\bar{E}$-vector is in the basal *ab*-plane, the bonding O $2p_{xy}$ and Ti $3d_{x^2\text{-}y^2}$, $3p_{xy}$, orbitals are probed along the $Ti_3C_2O_x$ surface, showing a peak at 5.2 eV as only $O_b$ is observed. When the $\bar{E}$-vector is perpendicular to the basal *ab*-plane (a), domains of $O_{fcc}$ are observed at a higher $E_b$ together with $O_b$ at 0.2 eV lower $E_b$.

The observations in Fig. 4 are supported by previous valence band XPS studies [24] that showed that for mixed $O_{fcc}$ and $O_b$ terminations, $O_{fcc}$ shifts to lower binding energy in the VB so that the peaks of $O_{fcc}$ and $O_b$ coincide. The peaks are well separated for horizontal polarization (Fig. 4(a)), indicating that there are areas on the $Ti_3C_2O_x$ surface where there is only $O_{fcc}$, while other areas have only $O_b$ or both $O_{fcc}$ and $O_b$. As shown by the polarization dependence in Fig. 4, the orbital extent and direction of the oxygen atom is different for $O_{fcc}$ and $O_b$, with one perpendicular and two angularly directed lone-pair orbitals, respectively. [23] (see Supplementary Information section S2).

The electrons at the FS control the transport properties, where the conductivity is proportional to the Fermi velocity ($v_F$), the concentration of Fermi electrons, *i.e.*, the density of states at $E_F$, and inversely proportional to the relaxation time [25]. Based on DFT calculations of the band dispersions and the effective mass approximation, the conductivity of $Ti_3C_2O_2$ is estimated to be ~ $1.8x10^5\Omega^{-1}m^{-1}$ with a relaxation time of about 1 fs [26] (see SI).

As the coverage of oxygen increases [24], there is an evolution of electronic properties as the adsorbed O experience mutual interaction with other O and alters the energy bands so that the bulk band gap within the occupied Ti *3d* states becomes narrower and shifts upwards in energy. At full oxygen coverage on the fcc-site of $Ti_3C_2O_x$, Bader charge calculations indicate that each adsorbed O gains an effective charge of -1.114e while the C atoms gain -1.667e and the central and surface Ti atoms loose +1.686e and +1.938e, respectively. The electronic orbitals and FS can be further modified by the coverage of the termination species on the bridge site between the Ti atoms. The 2D behavior of the electronic transport properties of $Ti_3C_2$-based MXene indicate that the interplanar interactions between the $Ti_3C_2O_x$ sheets are sufficiently weak so that the single sheet electronic features are highly preserved and behaves as stacked independent MXene sheets.

Different roles of the bands that crosses the $E_F$ both along and perpendicular to the basal plane are identified. The slope of the density of states (DOS) is rapidly increasing towards the conduction band. As the carrier velocities are perpendicular to the FS, they mainly contribute to the basal *ab*-plane except at the $\bar{\Gamma}$-point. O-terminated MXenes are predicted to show the highest resistivity among the inherent -F, -O and -OH species as the surface functionalization alters the electronic resistance through a decrease of the carrier concentration. Previous DFT studies predicted that termination of $Ti_3C_2T_x$ with -F, -O or -OH significantly alters and reduces the electronic states near $E_F$ [27,28], while a local maximum in the DOS at $E_F$ is predicted for bare $Ti_3C_2$. This enables that the in-plane electrical conductivity of the $Ti_3C_2T_x$ films can be tailored in a large range from zero to 2.9-4.8 $x10^5$ S/m [25], 2.4 $x10^5$ S/m [29] and 3.33 $x10^5$ S/m [30]. This is also supported by theoretical band structures, indicating that $Ti_3C_2O_2$ is metallic.

The calculated $v_F$ of $Ti_3C_2O_x$, with electron-like bands at the $\bar{\Gamma}$-point shows a velocity component along the $k_z$ axis. The other bands are active hole-like bands with large velocity contributions confined at the $\bar{K}$-point within the basal *xy*-plane (Fig. S1). These results thus show that $Ti_3C_2O_x$ is characterized by a clear carrier-type anisotropy, *i.e.*, positive (hole) charge carriers within the basal *ab*-plane and negative (electron) charge carriers govern the properties perpendicular to the surface. A parallel can be drawn with the thermoelectric properties (Seebeck coefficient) of the parent 3D MAX phases, where the electronic structure has been well studied [31]. Two types of bands contribute and sums up to the total Seebeck coefficient [29]; hole-like bands in the basal *ab*-plane and electron-like bands along the *c*- axis that both contribute but with opposite sign. This effect has been experimentally evidenced by anisotropic states in the electronic structure of $Ti_3SiC_2$ [32], which exhibits a highly unusual feature of near-zero Seebeck coefficient in a wide temperature range [33]. Similarly, $Ti_3C_2O_x$ behaves as a p-type material along the in-plane direction, having predominantly itinerant positive charges, while the behavior is n-type perpendicular to the basal planes. The type of termination species and coverage as well as the charge-carrier phonon coupling determines the carrier mobility of the material. This should also be true for related Ti-based MXenes such as $Ti_2CT_x$.

The results also have implications for intercalation of ions between 2D $Ti_3C_2O_x$ layers, a promising approach for energy storage devices, such as capacitors, batteries, and hydride electrochemical devices, where different ions can be considered as charge carriers [34]. $Ti_3C_2O_x$ could also be of significance as gas sensors of, *e.g.*, ethanol, methanol and acetone, $H_2O$, $H_2S$ and $CH_4$. $Ti_3C_2O_x$-based gas sensors should show high selectivity and sensitivity to adsorption of ammonia at room temperature [35]. This gas sensitivity mechanism is due to charge-transfer between the gas molecules and the O and Ti atoms on the surface. Upon adsorption, charge-transfer alters the shape of the FS and causes a significant change in the resistivity of $Ti_3C_2O_x$. This knowledge can be utilized for materials engineering of gas sensors, in energy storage units and water splitting for hydrogen production for circular electric energy production.

The development of such devices requires a proper description of the $Ti_3C_2O_x$ surface, including correct identification of the band structure, FS, termination species, surface site occupancies, and bonding configurations. This work provides information about the topology of the FS for pure O-termination yielding an exhaustive description of the fermiology, identifying conical band crossings with Dirac-like appearance at 1.5 eV, and brings us closer to a better understanding of the 2D properties of MXenes and their unique characteristics important for future energy storage devices, transparent conductors, and catalysis.

From a scientific perspective, conical band crossings below the $E_F$ are interesting because they may move in energy toward the $E_F$ by interaction with other states through surface reconstruction, hybridization or coupling effects, especially under doping or external field influence. Furthermore, the position of these states below the $E_F$ means they could participate in thermally activated processes, affecting the overall conductivity and potentially influencing phenomena like thermoelectric efficiency or magnetoresistance. Their stability and accessibility through techniques like ARPES make them important for investigating fundamental quantum effects in 2D materials and can guide efforts in band structure engineering for enhanced material performance in electronic applications.

Furthermore, the metallicity observed in oxygen-terminated $Ti_3C_2T_x$ MXene is intriguing in Fermiology because oxygen passivation generally leads to insulating surfaces by creating a bandgap. Typically, in surface science, oxygen termination of a material passivates surface states by saturating dangling bonds, effectively reducing the density of states at the Fermi level and shifting the material's behavior towards insulating. However, in $Ti_3C_2T_x$ MXene, despite oxygen termination, metallicity is preserved.

This behavior suggests unconventional electronic structure characteristics at the Fermi surface, with oxygen contributing electronic states that do not localize or create a significant bandgap. In the context of Fermiology, this is particularly compelling as it implies that the Fermi surface of these materials can host itinerant electrons despite surface passivation, potentially due to unique hybridization between oxygen and the Ti transition metal atoms. This phenomenon opens up new possibilities for studying electron transport, electron-phonon interactions, and potential superconductivity in materials with mixed metallic and insulating character.

**Conflict of interests:** The authors declare no competing interests.

**Acknowledgments:** We acknowledge the MAX IV Laboratory for the ARPES beamtime at BLOCH. Research conducted at MAX IV, a Swedish national user facility, is supported by the Swedish Research council under contract 2018-07152, the Swedish Governmental Agency for Innovation Systems under contract 2018-04969, and Formas under contract 2019-02496. The computations were enabled by resources provided by the Swedish National Infrastructure for Computing (SNIC) at the National Supercomputer Centre (NSC) partially funded by the Swedish Research Council through grant agreement no. 2016-07213. M.M. acknowledges financial support from the Swedish Energy Research (Grant No. 43606-1) and the Carl Tryggers Foundation (CTS23:2746, CTS20:272, CTS16:303, CTS14:310). The research leading to these results has received funding from the Swedish Government Strategic Research Area in Materials Science on Functional Materials at Linköping University (Faculty Grant SFO-Mat-LiU No. 2009-00971). P.E. also acknowledges funding from the Swedish Research Council (VR) under Project No. 2021-03826 and the Knut and Alice Wallenberg Foundation through the Wallenberg Academy Fellows program (grant no. KAW 2020.0196). We also thank Dr. Joseph Halim at Linköping University for preparing the thin film MAX phase through magnetron sputtering and the etching of the thin film sample.

**FIGURES**

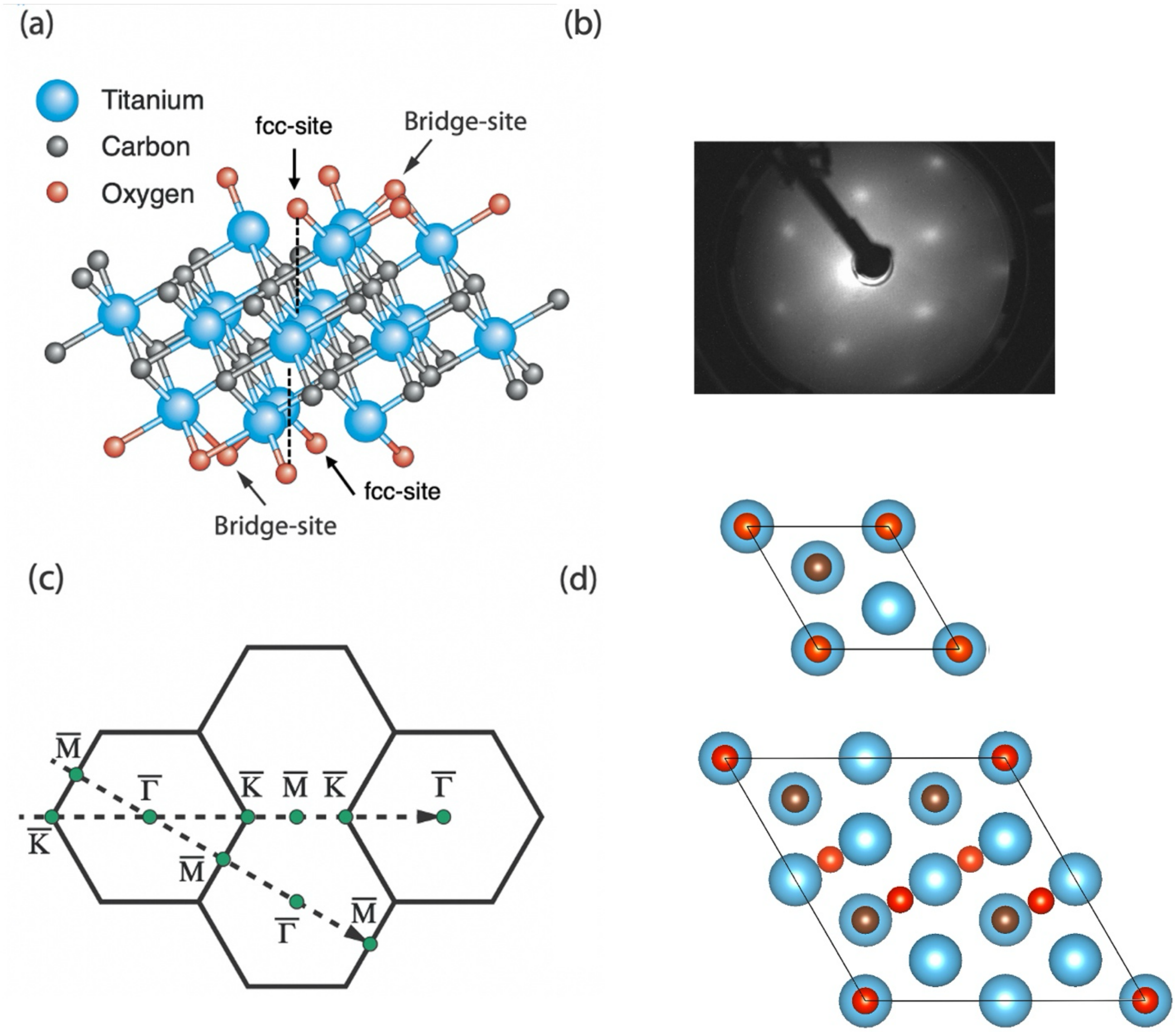

**Figure 1:** (Color online) (a) Crystal structure model of a $Ti_3C_2$ sheet terminated by O occupying hollow fcc-sites ($O_{fcc}$ between the surface Ti atoms on both upper and lower surfaces with optional bridge sites ($O_b$). (b) Representative LEED image after annealing showing hexagonal 1x1 spots taken at an electron beam energy of 98 eV. (c) Brillouin zones of the hexagonal honeycomb unit cell with symmetry points at: $\bar{\Gamma}(0,0,0)$ $\bar{M}$ $(\frac{1}{2}, 0, 0)$ $\bar{K}(\frac{2}{3}, \frac{1}{3}, 0)$. (d) Structure models of a $Ti_3C_2O_2$ 1x1 unit cell and a 2x2 $Ti_3C_2O_{1.5}$ supercell seen from above the surface.

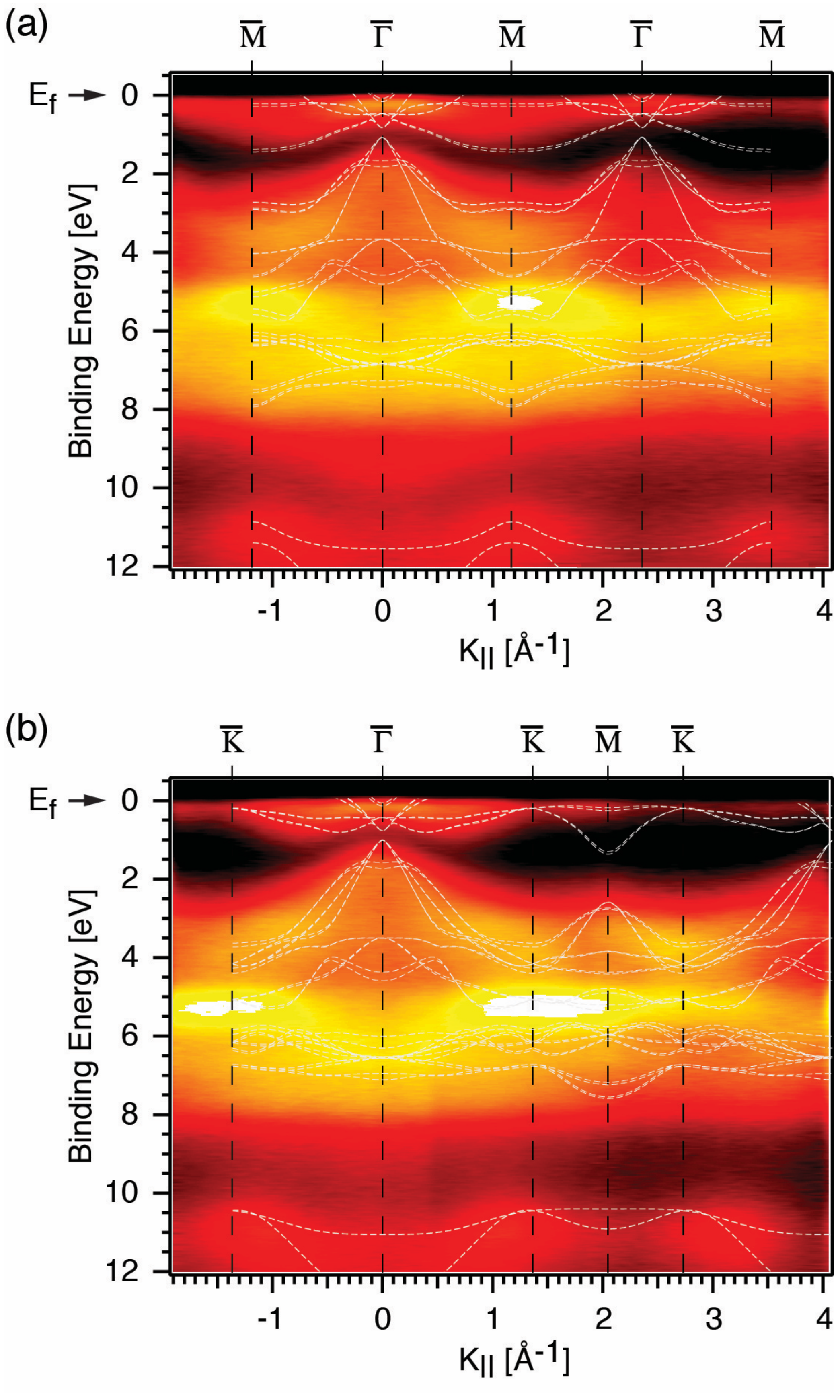

**Figure 2:** (Color online) (top) Band structure along the $\bar{M}$-$\bar{\Gamma}$-$\bar{M}$-$\bar{\Gamma}$-$\bar{M}$ high symmetry direction. (bottom) Band structure along the $\bar{K}$-$\bar{\Gamma}$-$\bar{K}$-$\bar{M}$-$\bar{K}$-$\bar{\Gamma}$ high symmetry direction. Black and white areas represent the lowest and the highest intensities, respectively, in a logarithmic intensity scale with an overlayed calculated band structure of $Ti_3C_2O_2$ unit cell model.

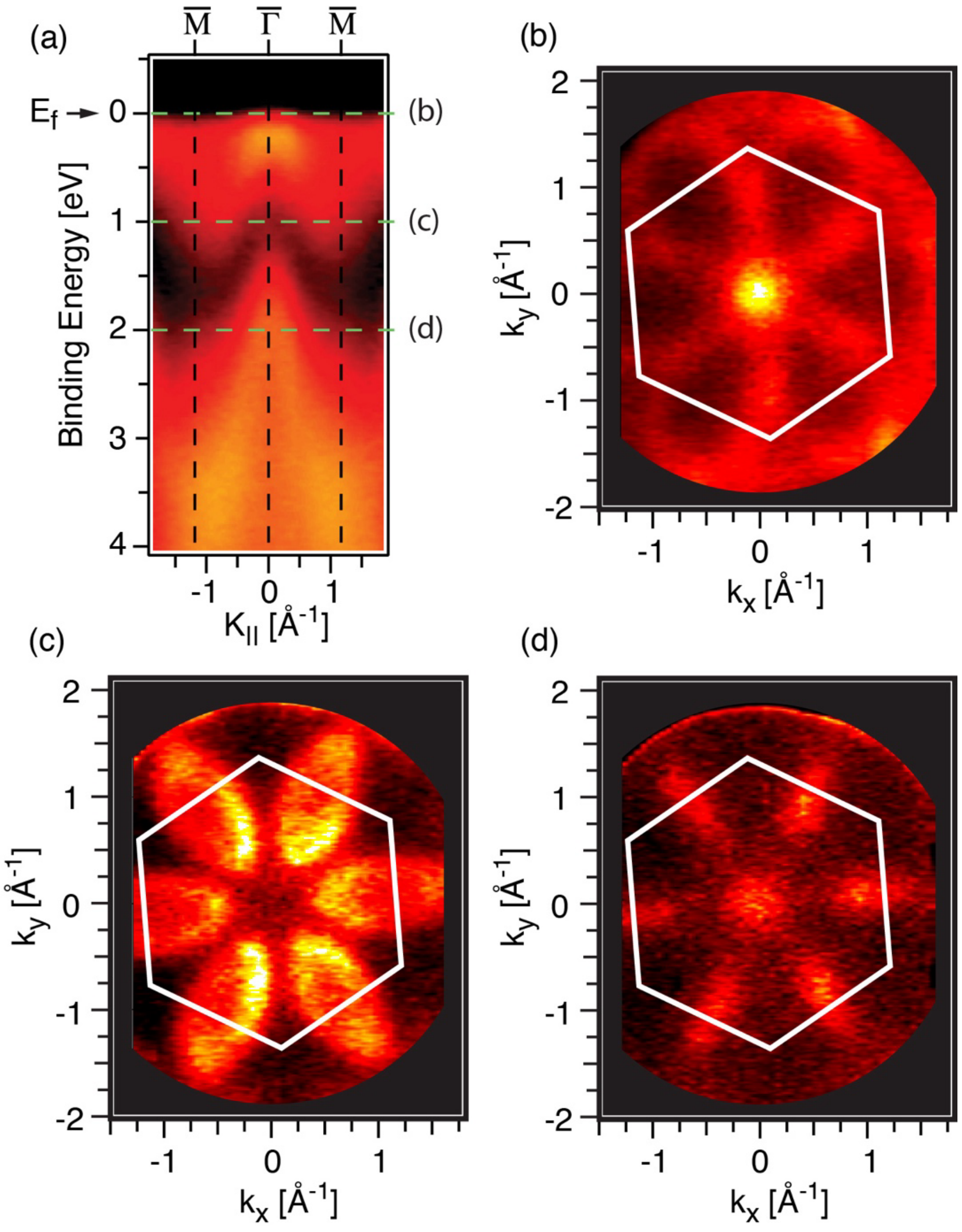

**Figure 3**: (Color online) (a) Measured band structure of $Ti_3C_2O_x$ at 180 eV. Constant energy contour surface maps at (b) the $E_F$ at 0 eV, (c) 1 eV, and (d) 2 eV below $E_F$ as indicated by the horizontal lines in (a).

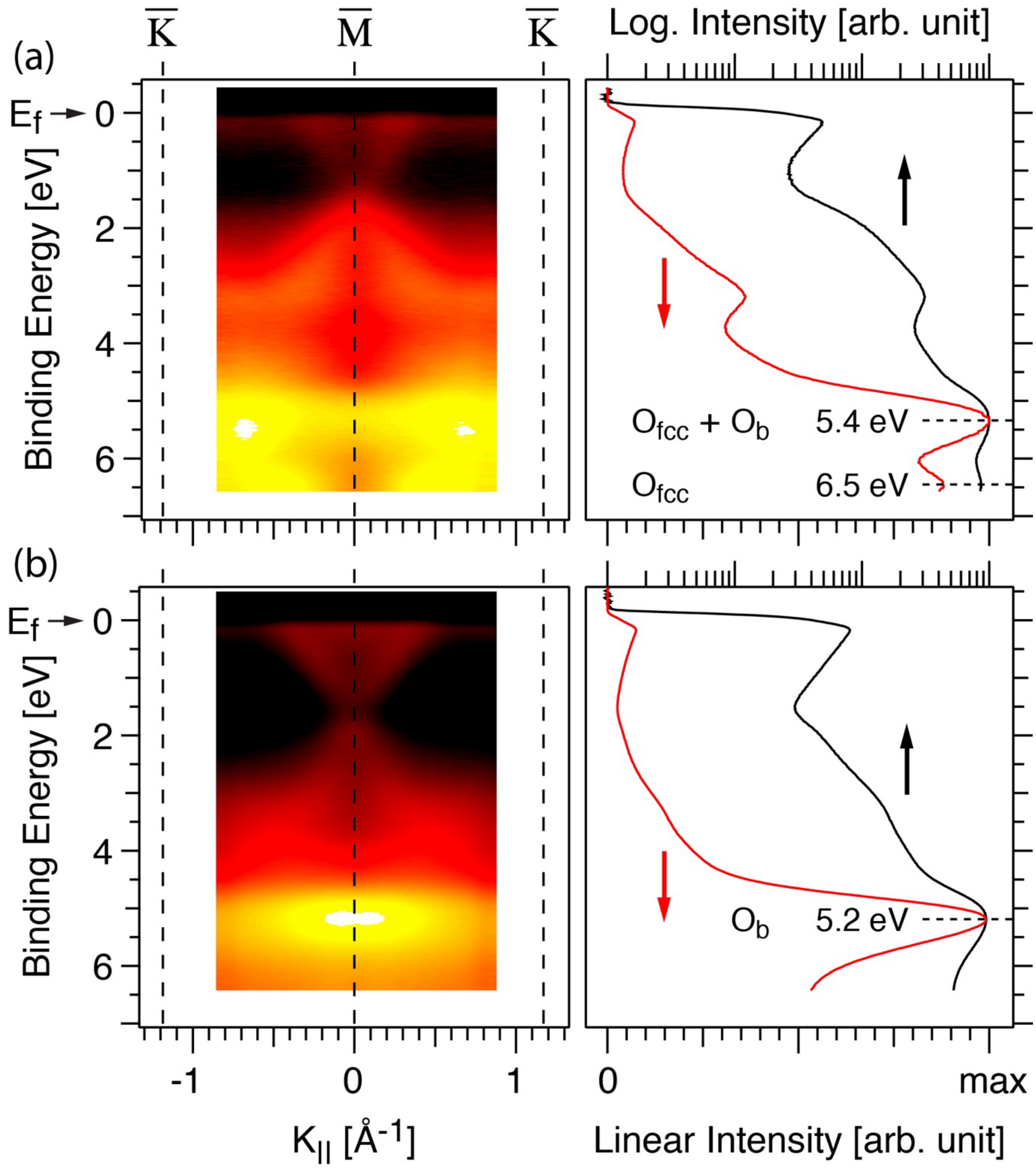

**Figure 4:** (Color online) Polarization-dependent measurements at 40 eV and 84 K along the BZ edge with integrated intensities in the right panels: (a) linear horizontal *s*-polarization with components both along the *c*-axis (O $2p_z$ and Ti $3d_{z^2}$ orbitals) and along the basal ab-plane. (b) linear vertical *p*-polarization along the basal *ab*-plane (O $2p_{xy}$ and Ti $3d_{x^2-y^2}$, Ti $2p_{xy}$ orbitals). The intense white areas around 5.2-5.4 eV on a logarithmic intensity scale (yellow color) are due to O $2p_z$ and O $2p_{xy}$ states, respectively, while the black areas represent the band gap.

# Supplementary information

# Fermiology and Band Structure of Oxygen-Terminated $Ti_3C_2T_x$ MXene

Martin Magnuson[1], Per Eklund[1,2], and Craig Polley[3]

*[1]Department of Physics, Chemistry and Biology (IFM), Linköping University, SE-581 83 Linköping, Sweden*
*[2]Inorganic Chemistry, Department of Chemistry - Ångström, Uppsala University, Box 538, SE-751 21 Uppsala, Sweden*
*[3]MAX IV Laboratory, Lund University, Fotongatan 2, SE-22484 Lund, Sweden.*

# S1. Methods

## S1.1. Sample preparation

MXenes synthesized as thin films are most suitable for ARPES measurements and these are the highet quality samples available. Synthesis of MXenes from thin MAX-phase films yields significantly more well-ordered, cleaner and atomically flatter surfaces with fewer impurities and dislocations, resulting in sharper, more intense XPS peaks compared to bulk and freestanding sheet/foil counterparts. The thin film $Ti_3C_2T_x$ stacked nanosheet sample was prepared from a thin film of $Ti_3AlC_2$ MAX-phase deposited on a *c*-axis-oriented sapphire ($Al_2O_3$) substrate with an area of 10x10 mm$^2$ using DC magnetron sputtering in an ultrahigh vacuum system. The deposition was performed using elemental Ti, Al and C targets with diameters of 75, 50 and 75 mm, respectively. Prior to deposition, the substrate was preheated inside the deposition chamber at 780 °C for 1 h. The Ti and C targets were ignited at 780 °C with powers of 92 and 142 W, respectively, for 30 s forming an incubation layer of TiC before the Al target was ignited at a power of 26 W. The duration of sputtering was 10 min, which produced a film of about 30 nm thickness. The $Ti_3AlC_2$ thin film was thereafter converted into $Ti_3C_2T_x$ MXene by etching in 50 % concentrated hydrofluorine acid for 1 h at room temperature. After etching, the sample was rinsed in deionized water and ethanol.

Formation of a high-quality MAX phase film was confirmed using Ni-filtered Cu Kα radiation (λ=1.5406 Å) in a normal Bragg–Brentano geometry of an X'Pert Panalytical x-ray diffraction (XRD) system [1]. Figure S1 shows ordinary θ-2θ diffractograms of the deposited $Ti_3AlC_2$ film and the etched $Ti_3C_2T_x$ MXene sample. Due to the limited diffraction volume, XRD of thin films exhibit higher noise levels compared to bulk materials and include peaks from both the substrate and growth layer. To minimize air exposure [2], the sample was subjected to XRD measurements for a brief duration of 10 minutes, after which it was stored under an argon (Ar) atmosphere. Mounting the $Ti_3C_2T_x$ film on the sample holder was performed in a glove-bag filled with nitrogen gas ($N_2$).

As observed in the diffractogram in Fig. S1, the main diffraction peaks originate from $Ti_3AC_2(000l)$ together with a smaller peak from the TiC(111) seed layer and a strong peak from the a-$Al_2O_3$(0006) substrate (S). The low intensities of the small additional peaks show that the films mainly consist of single-phase MAX phase. Furthermore, the fact that the diffractograms only show $Ti_3AlC_2$ of {000*l*}-type is a signature of epitaxial films [3]. The epitaxial growth behaviour has also been documented by transmission electron microscopy (TEM) [4].

The diffractograms clearly show the effect of HF etching on the Al layer. The position of the (002) peak, located at 2θ = 9.6893°, allows for the calculation of the interlayer spacing and c-axis=2*d002 by applying Bragg's Law. Based on the diffractogram presented in Fig. S1, the c-axis lattice parameter of the $Ti_3AlC_2$ film was determined to be 18.24 Å. After etching, the (002) peak shifts to a lower angle of 2θ = 6.5578°, corresponding to an expanded c-axis lattice parameter of 26.94 Å. The exact position of the (002) peak and the corresponding c-axis depends on the etching process [5,6].

The $Ti_3C_2T_x$ thin film sample on the $Al_2O_3$ substrate was stored in a vacuum suitcase and transported in Ar atmosphere to the MAX IV Laboratory synchrotron radiation facility and placed in the ARPES preparation chamber. Impurities such as $TiO_2$ and other surface contaminations were thereby avoided. The termination species ($T_x$) were identified in a previous study as fluorine (F) and oxygen (O) by core-level X-ray photoelectron spectroscopy (XPS) [7]. A heat treatment up to 750 °C for 10 hours removed all the F species and pure O termination was obtained.

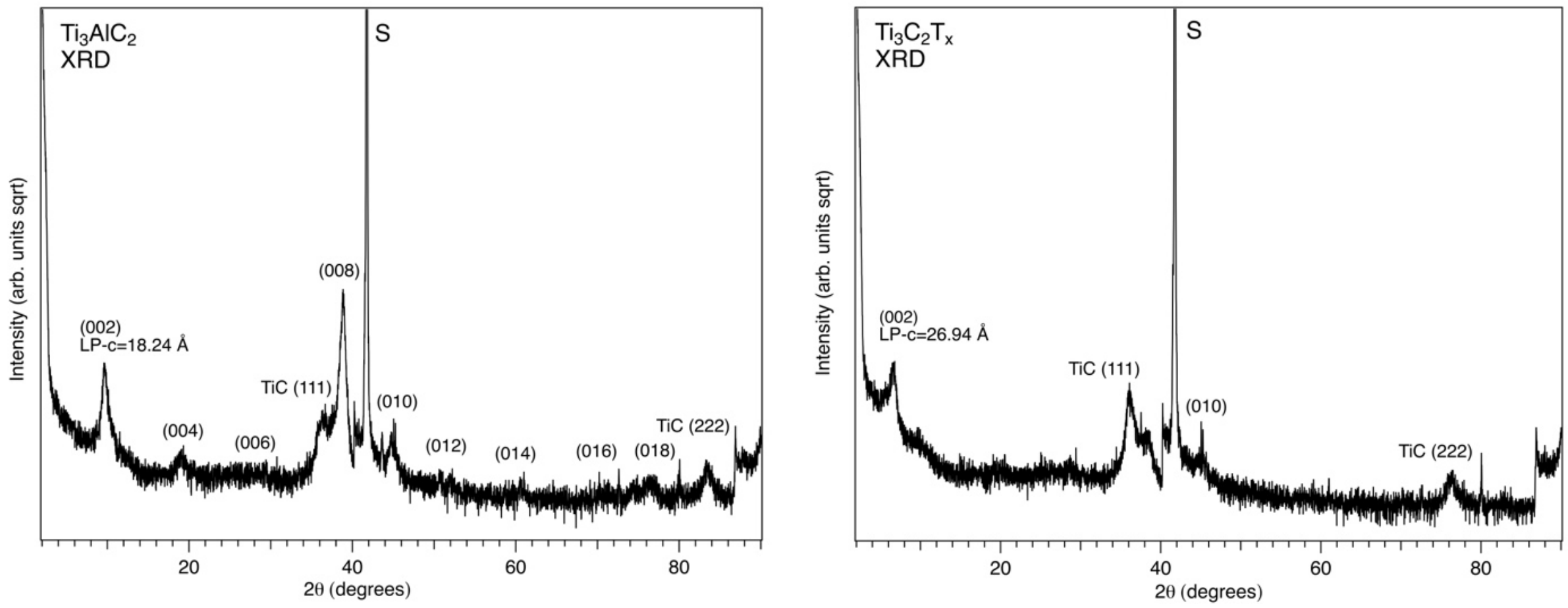

**Figure S1:** X-ray diffractograms (2-90°) of $Ti_3AlC_2$ MAX-phase and $Ti_3C_2T_x$ MXene, where the (002) peak is located at 2θ=9.689° and 6.558°, respectively.

## S1.2. ARPES measurements

The ARPES data were measured at the BLOCH beamline at the 1.5 GeV storage ring at the MAX IV Laboratory in Lund, Sweden. Initially, degassing was applied using a low direct current through the sample in UHV ($4e^{-10}$ mBar) for 5 hours. Two different annealing steps were then applied. In the first annealing step with 0.3A/13.9V, the sample gently glowed dull red >500°C for 10 minutes at a pressure of $1.9e^{-9}$ mBar. In the second annealing step, the current was increased to 0.4 A for 5 minutes at >600-700 °C. After the second step, the LEED image shown in Fig. 1b was taken at a beam energy of 98 V and a filament current of 1.85 A in the ARPES chamber.

During the ARPES measurements, the spot size on the sample surface was 10 × 10 μm. A Scienta Omicron DA30-L hemispherical electron energy analyzer was used for electron detection. Overview ARPES data were obtained using a photon energy of 180 eV at 45 K with a pass energy of 50 eV. Due to a small amount of charging of the underlying sapphire substrate, each spectrum was slightly shifted to bring the intensity cutoff at the sample Fermi level ($E_F$) to a binding energy of zero. The small photon-induced charging offset varied with photon flux and energy but remained otherwise stable over time. Further ARPES measurements using horizontal s- and vertical p-polarization at low photon energy (40 eV) were performed to gain insight about 2D features and the orbital character. Qualitatively, the bands were the same everywhere on the sample but the O coverage appeared to be different in certain spots on the surface.

## S1.3. Computational details

First-principles calculations were performed by means of density functional theory (DFT) and the projector augmented wave method (PAW) [8] as implemented in the Vienna *ab-initio* simulation package (VASP) [9]. The PAW method was used together with the exchange-correlation (XC) functional PBE-SOL [10] within the Generalized Gradient Approximation (GGA), incorporating the Grimme van der Waals DFT-D2 scheme [11] and accounting for the effect of spin-orbit interaction. The plane-wave expansion cutoff energy was set to 700 eV, with 80 k-points between each pair of high-symmetry points and a 30x30x30 k-point mesh for the Fermi surface calculations.

Calculations we made using a 1x1 primitive cell $Ti_3C_2O_2$ multilayer structure with O located at the threefold fcc site between three surface Ti atoms with a unit cell width of 3.075 Å and a height of 20.51 Å. The Ti-O distance was 2.047 Å. GGA+U with U=2.8 eV and a PBE-SOL potential was applied to improve the description of the Ti 3d electrons. The on-site Coulomb correction (GGA+U) [12] for Ti was calculated as U=2.8 [13], reflecting the screened atomic correlation energy [14].

A 2x2 supercell with a bridge-site coverage of $Ti_3C_2O_{0.25}$ on each surface side was needed to explain a faint flat band along the $\bar{\Gamma}$-$\bar{M}$ direction that suggests a party mixed phase of termination species sites. A representative 2x2 $Ti_3C_2O_1O_{0.5}$ supercell with mixed oxygen sites (total x=1.5) used in the calculations is shown in Fig. 1(d) including oxygen on both fcc and bridge sites. The bands in the 2x2 supercell were unfolded into a 1x1 primitive cell representation by applying the effective band structure method using the program BandUP [15]. The Fermi velocities of $Ti_3C_2O_2$ and $Ti_3C_2$ were calculated using the FermiSurfer software [16].

## S2. Fermi velocities

Figure S2 shows calculated Fermi velocities for a) $Ti_3C_2O_2$ and b) $Ti_3C_2$ (bare MXene) of multilayer structures with electron and hole bands crossing the $E_F$ and c) closeups of the active bands centred around the $\bar{M}$ symmetry point with electron bands in the corners at the $\bar{\Gamma}$ symmetry point. The carrier velocities are orthogonal to the FS and are reflected the inclination of the bands where they cross the $E_F$. A higher inclination implies higher Fermi velocity of the charge carriers and increases the transport properties. The values of the Fermi velocities were deduced from the slope of the highest band dispersion that crosses the $E_F$ around the $\bar{K}$-points. As shown in Fig. S2a) for $Ti_3C_2O_2$, the Fermi velocity is highest in the red-colored bands in the vicinity of the $\bar{K}$-points in the basal *ab*-plane while there is mostly an out of the plane velocity component at the $\bar{\Gamma}$-point. As shown for metallic bare MXene in Fig. S2b), the plaquettes of the FS should give rise to more out-of-plane conduction than in the case of $Ti_3C_2O_2$ that has a narrow rod-like surface state feature at the $\bar{\Gamma}$-point. The transversal parts of the FS in $Ti_3C_2O_2$ results in significant in-plane conduction. At zero temperature, the bands up to the FS corresponds to occupied (ground) states while those above are unoccupied (excited) states. The occupied states close the FS determines the plasmon frequencies of the electron and hole carriers as well as the electrical conductivity. As the electron velocities are orthogonal to the FS, the flat hole bands wrapped around the $\overline{\mathrm{K}}$-points are responsible for the in-plane conductivity in $Ti_3C_2O_2$, indicating that the conductivity is much higher in the basal *ab*-plane than along the *c*-axis. This can also be corroborated with the large difference in effective masses of the holes and electrons in-plane vs out-of-plane that gives rise to a significant anisotropy in the conduction properties. In the basal *ab*-plane, the effective masses of the holes are an order of magnitude smaller than those of the electrons that leads to and increased carrier mobility [14]. This shows that mainly hole carriers are responsible for the in-plane conduction in $Ti_3C_2O_2$.

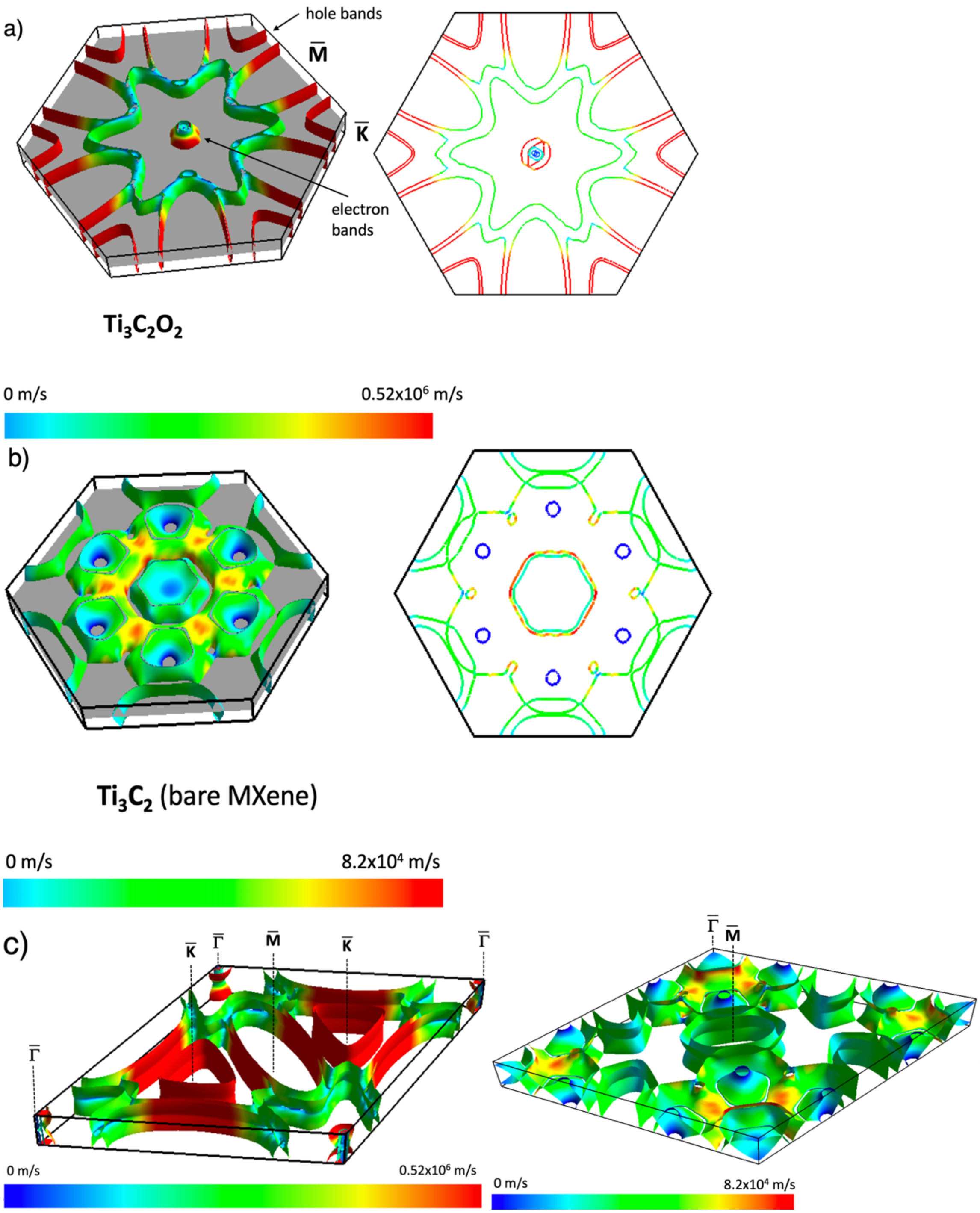

**Figure S2:** a) (left) Calculated Fermi velocities $v_F$ for a $Ti_3C_2O_2$ multilayer structure compared to b) $Ti_3C_2$ (bare MXene) with electron and hole bands crossing the $E_F$. Fermi lines in the $\bar{\Gamma}$-plane in the Brillouin zone is shown in the right figures. c) Closeup of the active hole bands (red) in $Ti_3C_2O_2$ (left) compared to less active bands in $Ti_3C_2$ (right) in rhombohedral Brillouin zones centered around the $\bar{M}$ symmetry point. The FS velocity moduli $v_F$ is shown by the color scales.

## S3. Charge-density difference plots

Figure S3 illustrates a tentative bonding model shown as calculated charge-density difference plots of $Ti_3C_2O_x$ surfaces with -O (red) adsorbed on a three-fold coordinated hollow fcc-site (a) involving σ-bonds in a yellow triangular negatively charged volume, where the corners are directed towards the three Ti atoms with positively charged volumes in blue and (b): on a bridge site between two Ti atoms with π-bonds the O atoms obtain a linear yellow negatively charged volume. The direction of the electronic charge-transfer is indicated by arrows that is also similar in the case of additionally adsorbed $H_2O$ on top of the O species.

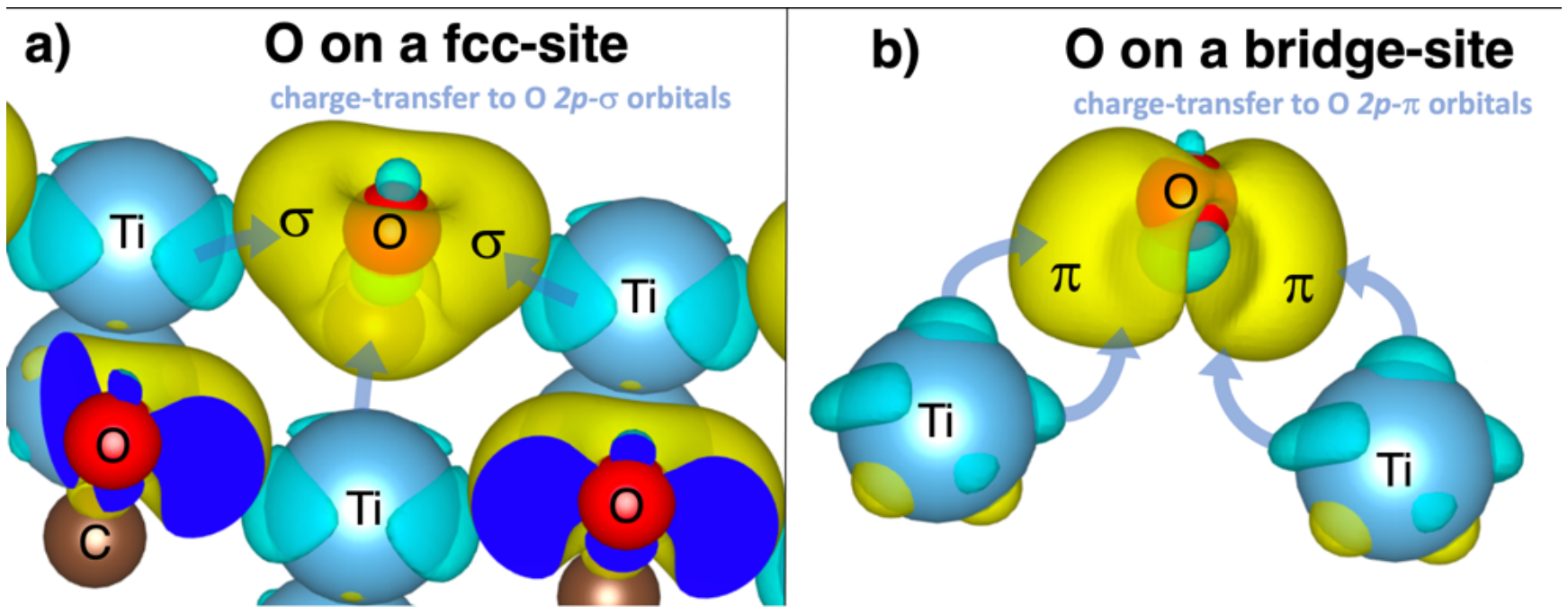

**Figure S3:** Charge-density difference plots of $Ti_3C_2O_x$ surfaces with -O (red) adsorbed on a three-fold hollow fcc-site between three Ti atoms gives σ-bonds (b) Bridge site between two Ti atoms gives π-bonds. Blue and brown spheres are Ti and C atoms and the arrows indicate charge-transfer from Ti toward the O atoms.

## S3. Band structure details

Figure S4 shows a close-up of the bands along the $\overline{\Gamma}$-$\overline{M}$-$\overline{K}$-$\overline{\Gamma}$ high symmetry E-k cuts extracted from a 3D map measured with horizontal polarization at 180 eV and 45 K. The dispersive electron and hole bands in the vicinity of the $E_F$ are responsible for the formation of electron and hole pockets in the vicinity of the FS. Notably, the bands at the $\overline{\Gamma}$-point and $\overline{M}$-points form electron pockets mainly consisting of $d_{z2}$ orbitals perpendicular to the basal plane while the bands at the $\overline{K}$-points at the corners of the BZ form hole pockets that mainly consists of in-plane $d_{xy}$ and $d_{x2-y2}$ orbitals. As a result of the strong Ti *3d* – C *2p* covalent bonding in the basal plane, the energy dispersion of the band structure is much stronger in-plane than along the *c*-axis giving rise to a characteristic 2D character of the material. Thus, the metallic conduction predominantly originates from the Ti *3d* states in the vicinity of the $E_F$. Generally, the band structure of $Ti_3C_2T_x$ exhibits similarities to that of graphene at the $\overline{K}$ and $\overline{M}$ points. This is attributed to the behavior of bulk MXene as an assemblage of independent single MXene layers, analogous to the structure of multilayer graphene. When including the spin-orbit interaction in the modelling, two quasi-double-degenerated bands occur with a splitting of 12.9 meV at the $\overline{K}$-point.

The experimental band broadening compared to theoretical predictions is attributed to the $k_z$-dispersion (k-orthogonal), as comprehensively described in the references [18,19,20]. In three-dimensional (bulk) materials, ARPES reveals a strong and well-defined E(k) dependence. In contrast, for a monolayer, such as graphene, there is minimal k-dependence, resulting in a nearly straight line at a specific energy. When ARPES probes more than one monolayer (2-3 monolayers, with a probe depth of approximately 5 Å at 50 eV), the contributions from layers beyond the top atomic layer shift the energies into a curvature, leading to a broadening of E(k). This broadening has been observed in ARPES measurements of $Ti_3SiC_2$ by Pinek *et al.* [21], who also investigated $V_2AlC$ [22], $Ti_2SnC$ [23], and $Cr_2AlC$ [24]. To measure the band structure and FS, Schultz *et al.* [25] applied ARPES on delaminated $Ti_3C_2T_x$, but the terminations were mixed, and the data was azimuthally averaged due to randomly oriented flakes in the measured film, hindering measurements along specific symmetry lines and the extraction of the shape of the FS.

Figure S5 shows photon energy scans (60-180 eV) at 45 K along the $\overline{M}$-$\overline{\Gamma}$-$\overline{M}$ and $\overline{K}$-$\overline{M}$-$\overline{K}$ directions with the energy independent features (dark areas) along the horizontal dashed lines close to the $E_F$. These energy-independent features are hallmarks of 2D surface states because they reflect the lack of dispersion with momentum perpendicular to the surface, indicative of states localized at the surface layer. These features provide crucial insights into the electronic structure and dimensionality of the material being studied.

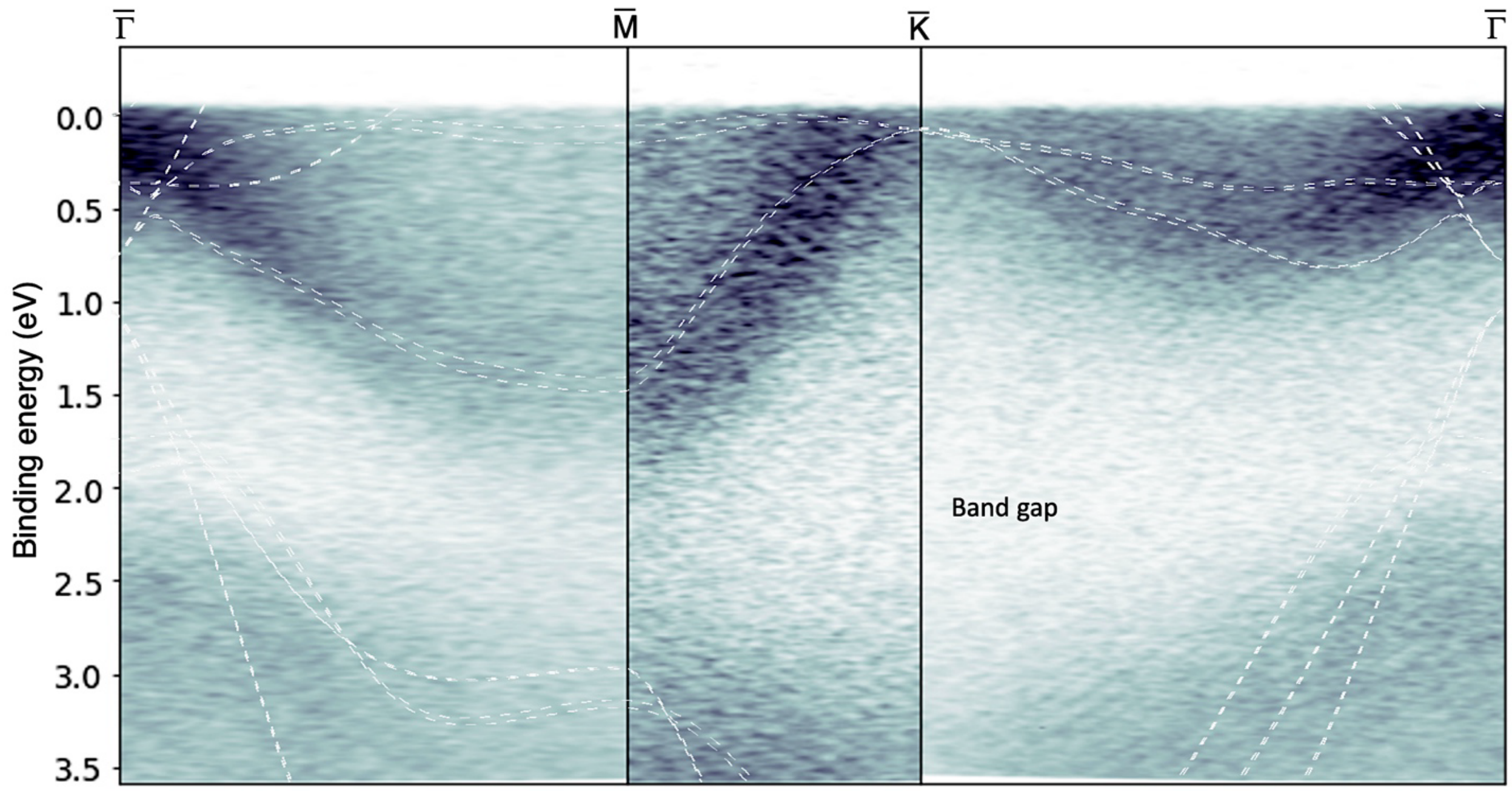

**Figure S4:** Close-up of a $\overline{\Gamma}$-$\overline{M}$-$\overline{K}$-$\overline{\Gamma}$ high symmetry cut extracted from a 3D map measured with horizontal polarization at 45 K. Calculated bands are overlayed in white color.

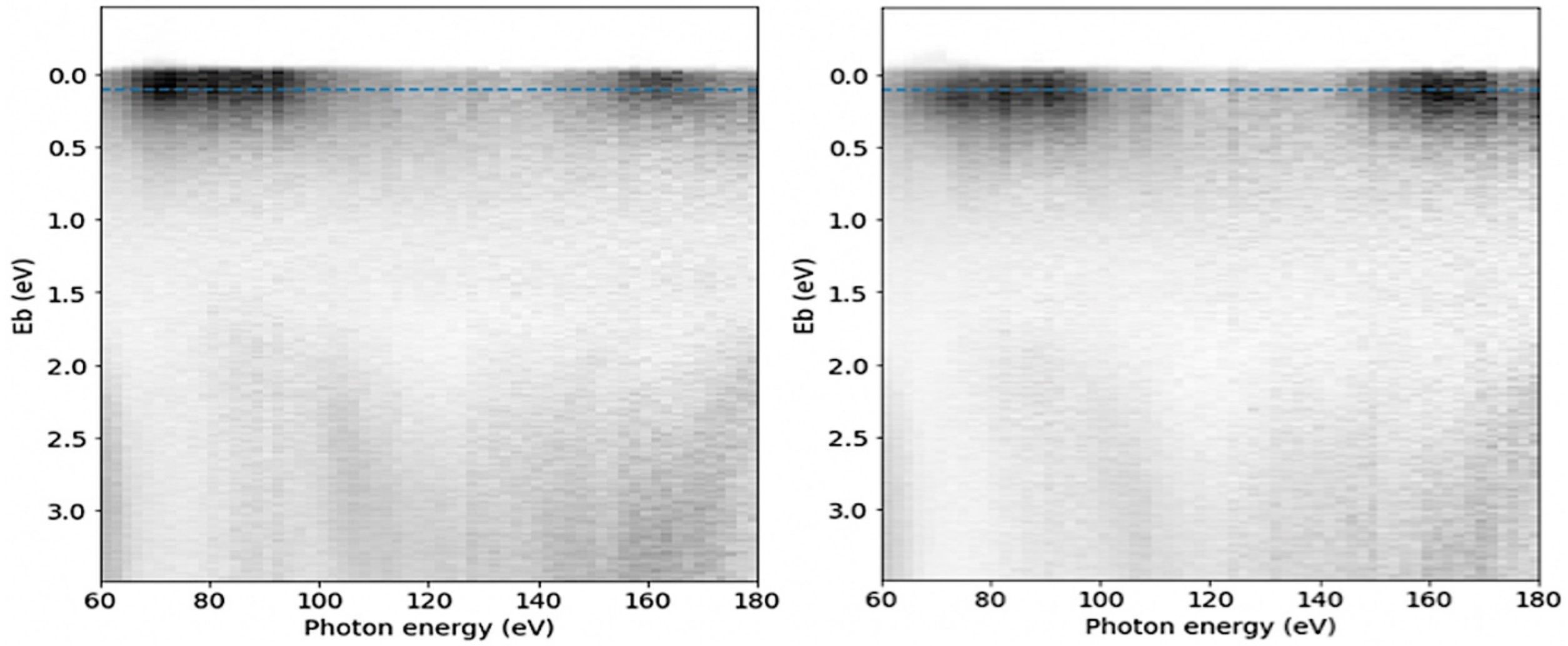

**Figure S5:** Photon energy scans (60-180 eV) at 45 K along the $\overline{M}$-$\overline{\Gamma}$-$\overline{M}$ and $\overline{K}$-$\overline{M}$-$\overline{K}$ directions where the energy independence indicates 2D surface states along the horizontal dashed lines close to the $E_F$.

Figure S6 shows core-level XPS spectra measured at the same beamtime as the LEED and ARPES experiments. The overview spectrum (a) shows the expected peaks, and that the surface is clean from contaminats and only contains Ti, C and O. The Auger peaks in the overview spectrum were identified by altering the photon energy, as their energy positions remain constant on a kinetic energy scale but shift on a binding energy scale, unlike the core-level peaks.

The O *1s* spectrum (b) exhibits a double feature where the peak at 531.2 eV is due to O bonded to the fcc site ($O_{fcc}$) while the peak at 530 eV is due to O bonded to the bridge site [26]. Initially, before annealing, regions of the sample only showed the $O_{fcc}$ peak at 531.2 eV. Previous combined XPS and TEM studies also showed that O occupies two different sites on the $Ti_3C_2$ surface [27]. The O component with the highest $E_b$ and largest bandwidth corresponds to $O_{fcc}$ with chemical bonds to three Ti surface atoms in a hollow site while the $O_b$ component with the lowest $E_b$ and smallest bandwidth corresponds to a bridge site between two surface Ti atoms [26]. A previous valence band XPS study [28] showed that for mixed $O_{fcc}$ and $O_b$ terminations, $O_{fcc}$ shifts to lower binding energy in the VB so that the peaks of $O_{fcc}$ and $O_b$ coincide. In case there is only $O_{fcc}$ or if there is only $O_b$ then these peaks are separated.

The Ti *2p* spectrum (c) shows that the sample surface is free from oxidized material such as $TiO_2$, as there are no oxide peaks at approximately 459 eV and 465 eV [29-31]. The single C *1s* peak (d) at 282.0 eV shows that the sample surface is free from C contamination. Contaminated MXene surfaces exhibit a peak at approximately 284.5 eV, due to graphite-like carbon, a residual product from etching, while hydrocarbon groups produce a peak at 284.5 eV [26,27,29].

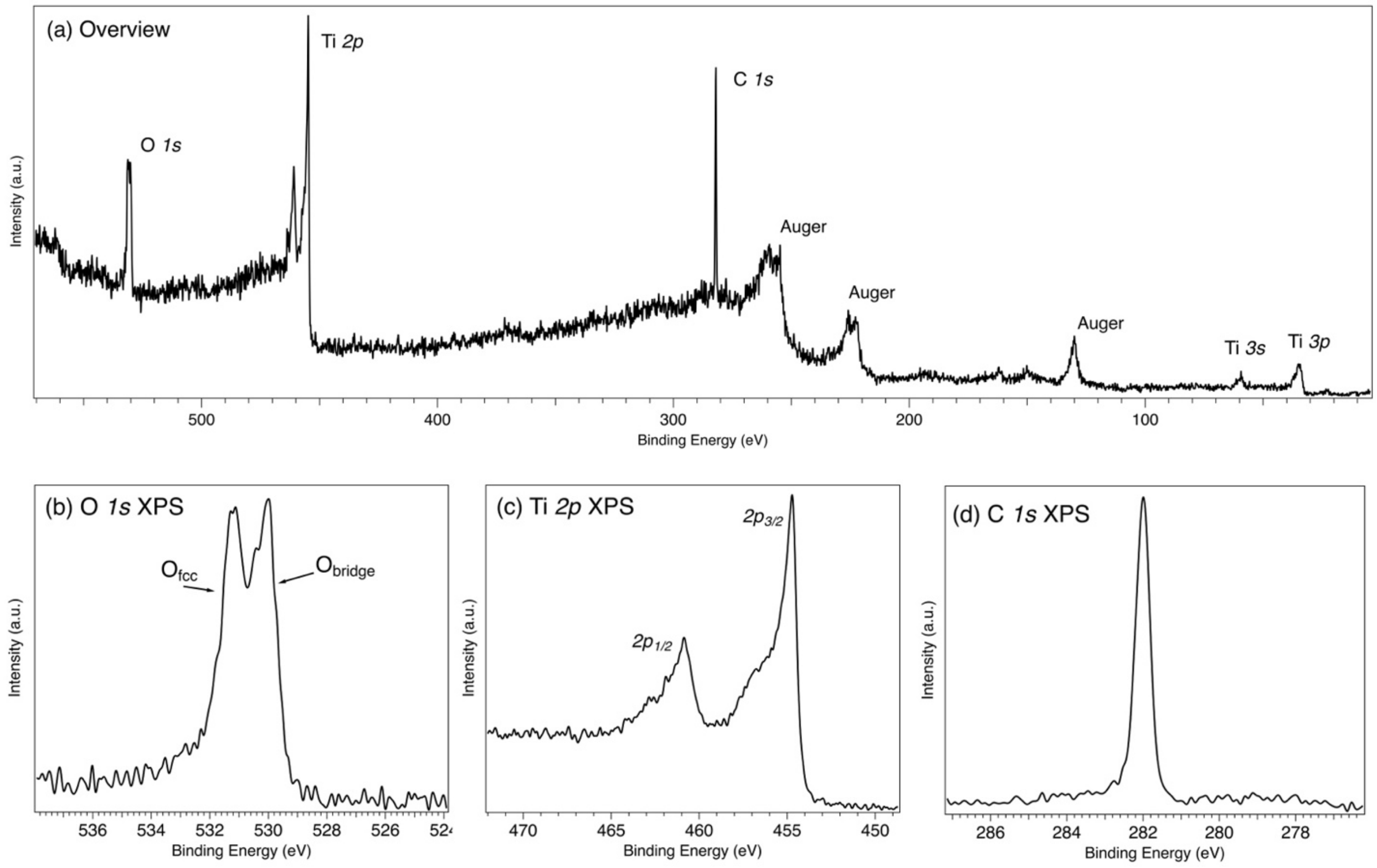

**Figure S6:** (top) overview XPS spectra of $Ti_3C_2O_x$ and (bottom) core level XPS (a) O *1s*, (b) Ti *2p*, and (c) C *1s*. The O *1s* spectra show that the $Ti_3C_2O_x$ sample has O surface termination both on the fcc-site and on the bridge-site. All spectra were measured with a photon energy of 642 eV.

## S4. Coverage-dependent flat band along the $\overline{M}$-$\overline{\Gamma}$-$\overline{M}$ direction

Figure S7: (left) shows detailed E-k cuts measured at 120 eV with horizonal polarization at 20 K along the $\overline{M}$-$\overline{\Gamma}$-$\overline{M}$ direction. The second derivative (right) reveals a weak flat band at 1.45 eV. Black and white areas represent the highest and lowest intensities, respectively, on a linear intensity scale.

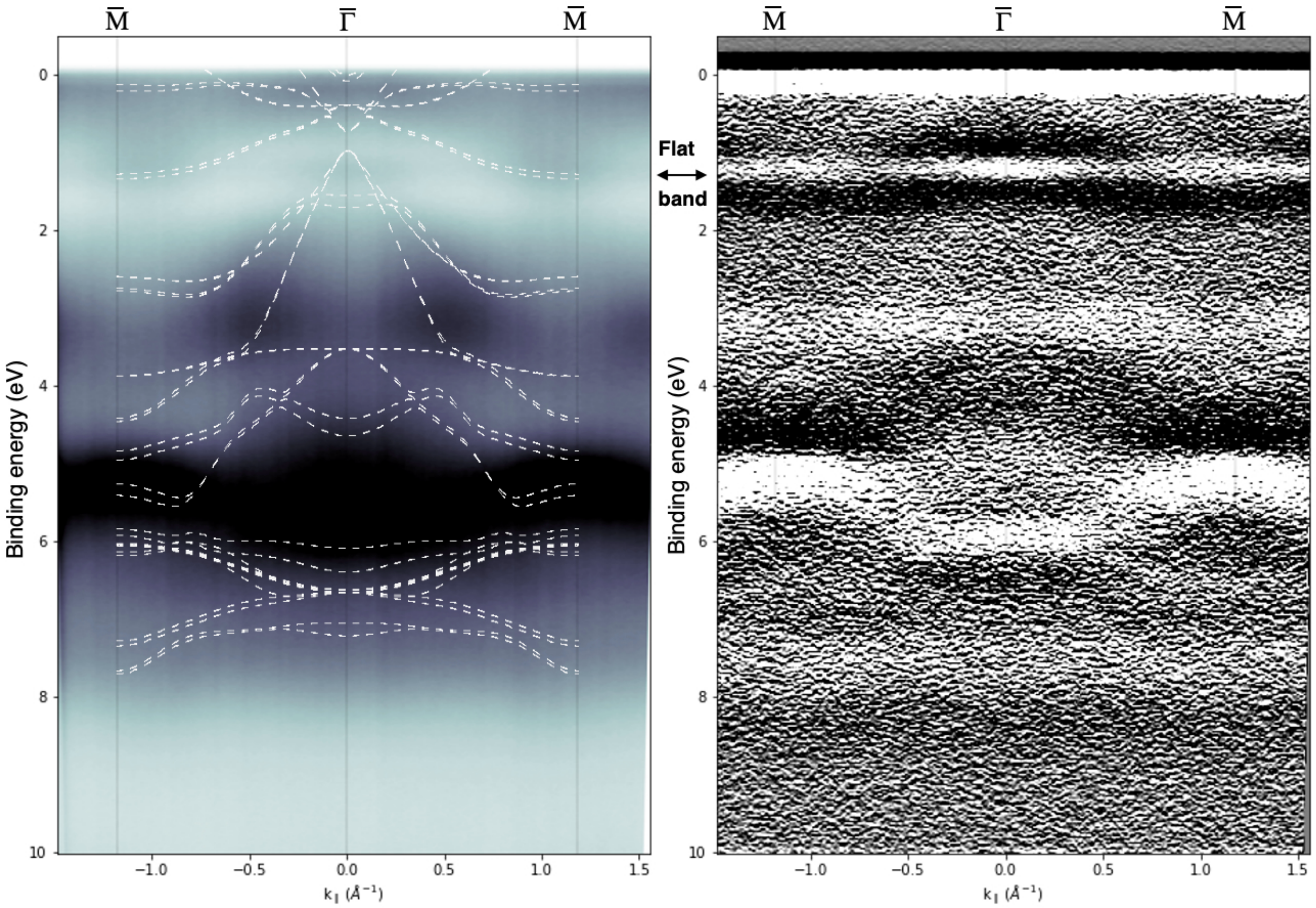

**Figure S7:** (left) Detailed E-k cuts measured at 120 eV with horizonal polarization at 20 K along the $\overline{M}$-$\overline{\Gamma}$-$\overline{M}$ direction. The second derivative (right) reveals a weak flat band at 1.45 eV. Black and white areas represent the highest and lowest intensities, respectively, on a linear intensity scale.

Figure S8a-d: shows a set of more detailed E-k cuts measured at 120 eV with horizonal polarization at 20 K along the $\overline{M}$-$\overline{\Gamma}$-$\overline{M}$ and $\overline{K}$-$\overline{M}$-$\overline{K}$ directions. The second derivative on the right reveals a weak flat band at 1.5 eV along the $\overline{M}$-$\overline{\Gamma}$-$\overline{M}$ direction and a Dirac-like band crossing along the $\overline{K}$-$\overline{M}$-$\overline{K}$ direction.

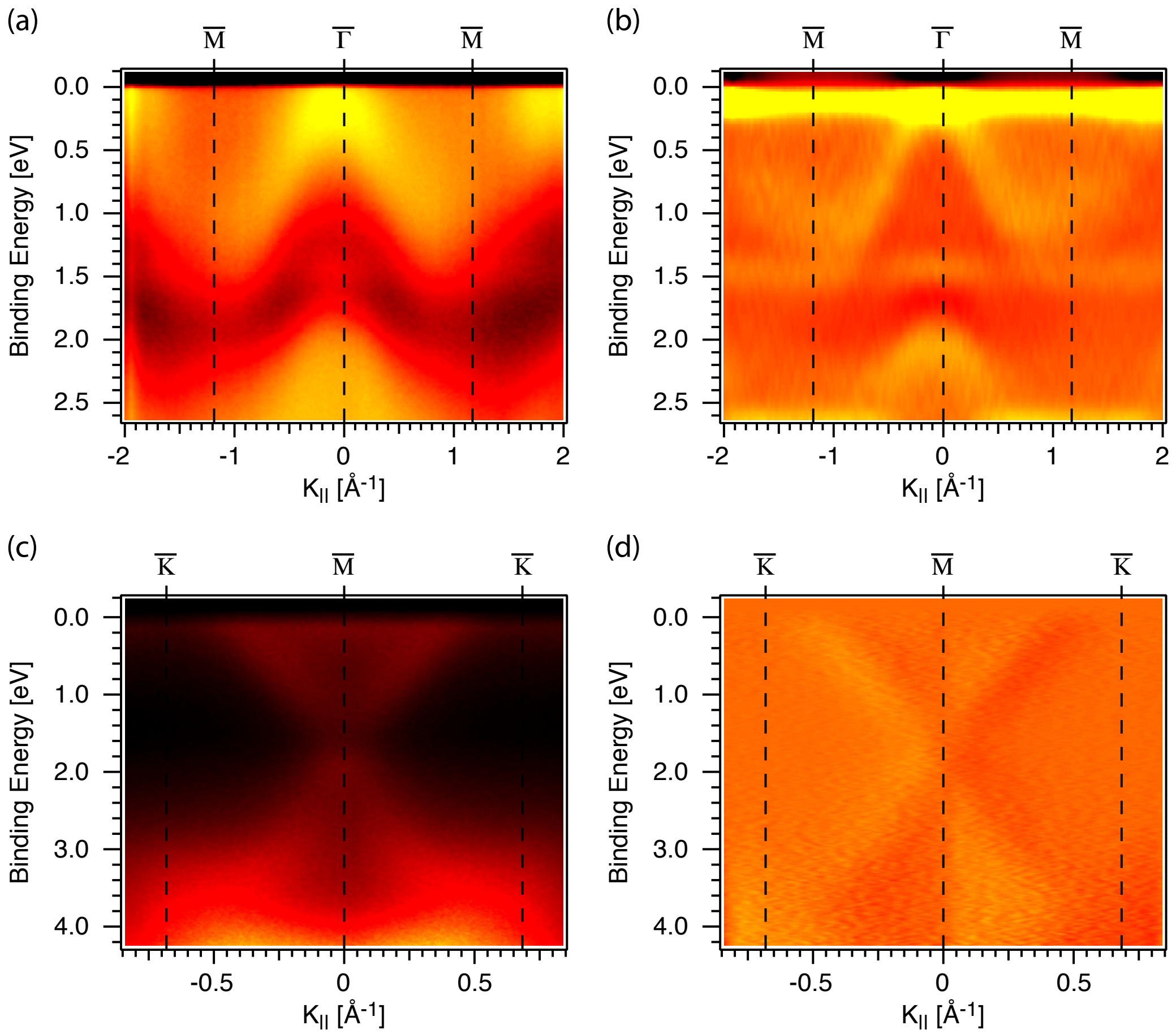

**Figure S8:** (a, c) Detailed E-k cuts measured at 120 eV with horizonal polarization at 20 K along the $\overline{\mathrm{M}}$-$\overline{\Gamma}$-$\overline{\mathrm{M}}$ and $\overline{\mathrm{K}}$-$\overline{\mathrm{M}}$-$\overline{\mathrm{K}}$ directions revealing a Dirac-like band crossing at 1.55 eV at the $\overline{\mathrm{M}}$-point. The second derivative (b, d) on the right reveals a weak flat band at 1.45 eV along the $\overline{\mathrm{M}}$-$\overline{\Gamma}$-$\overline{\mathrm{M}}$ direction. Black and yellow areas represent the lowest and highest intensities, respectively, on a linear intensity scale.

## S5. Unfolded supercell band structure calculations

Figure S9 shows unfolded bands along the $\overline{K}$-$\overline{M}$-$\overline{K}$ direction of a $Ti_3C_2O_{fcc}$ supercell model exhibiting a Dirac-like band crossing at 1.5 eV at the $\overline{M}$-symmetry point.

Figure S10 shows unfolded bands along the $\overline{M}$-$\overline{\Gamma}$-$\overline{M}$-$\overline{\Gamma}$-$\overline{M}$ and $\overline{K}$-$\overline{\Gamma}$-$\overline{K}$-$\overline{M}$-$\overline{K}$-$\overline{\Gamma}$ directions of $Ti_3C_2O_{fcc}O_b$ in a 2x2 supercell model (right). A faint flat band is discerned at 1.5 eV only along the $\overline{M}$-$\overline{\Gamma}$-$\overline{M}$ direction in some areas of the sample. Mixed $O_{fcc}$ and $O_b$ oxygen sites were needed to reproduce this band.

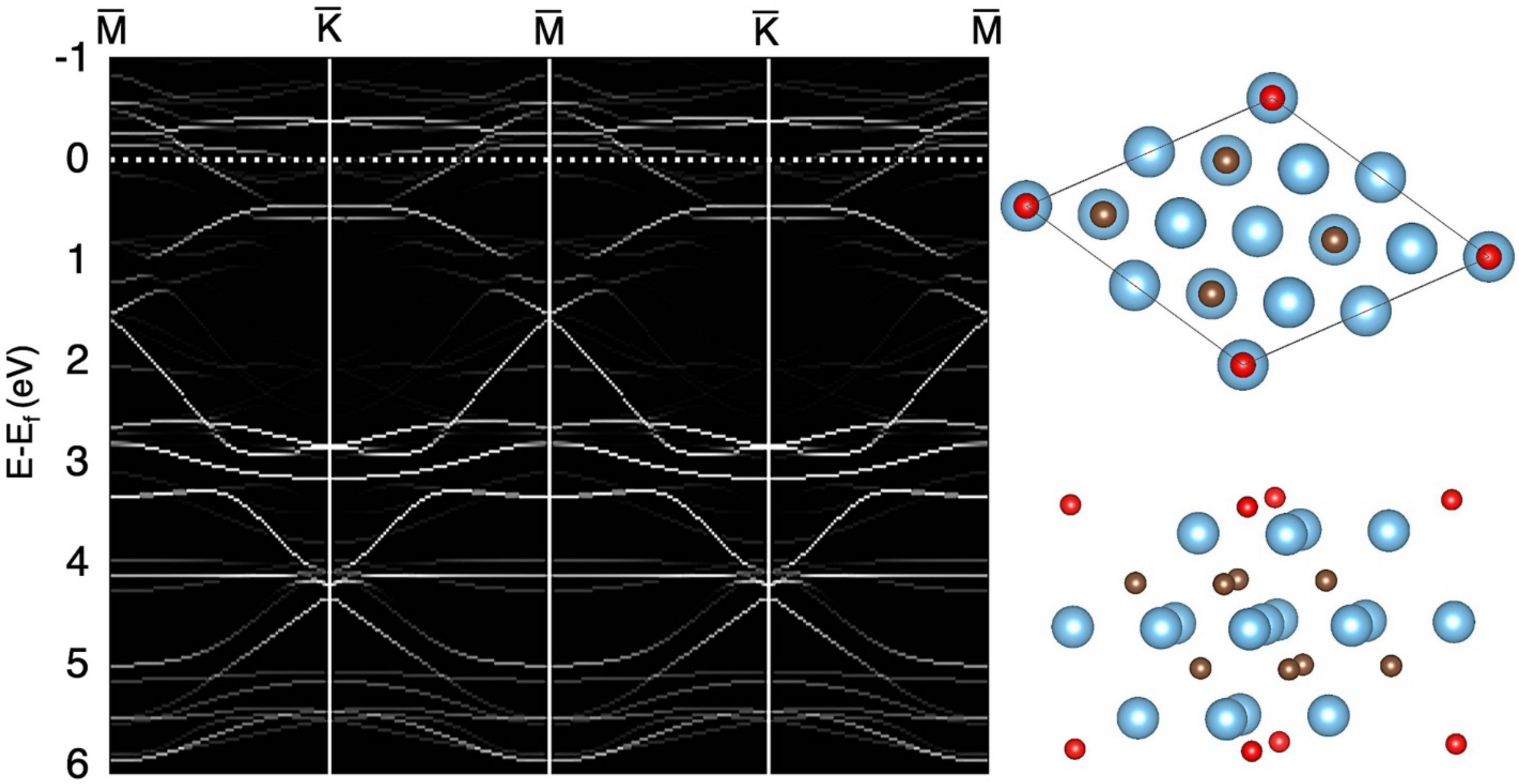

**Figure S9:** Unfolded bands along the $\overline{K}$-$\overline{M}$-$\overline{K}$ direction of $Ti_3C_2O_{fcc}$ (left) with a 2x2 supercell model (right) shows a band crossing at 1.5 eV at the $\overline{M}$-symmetry point.

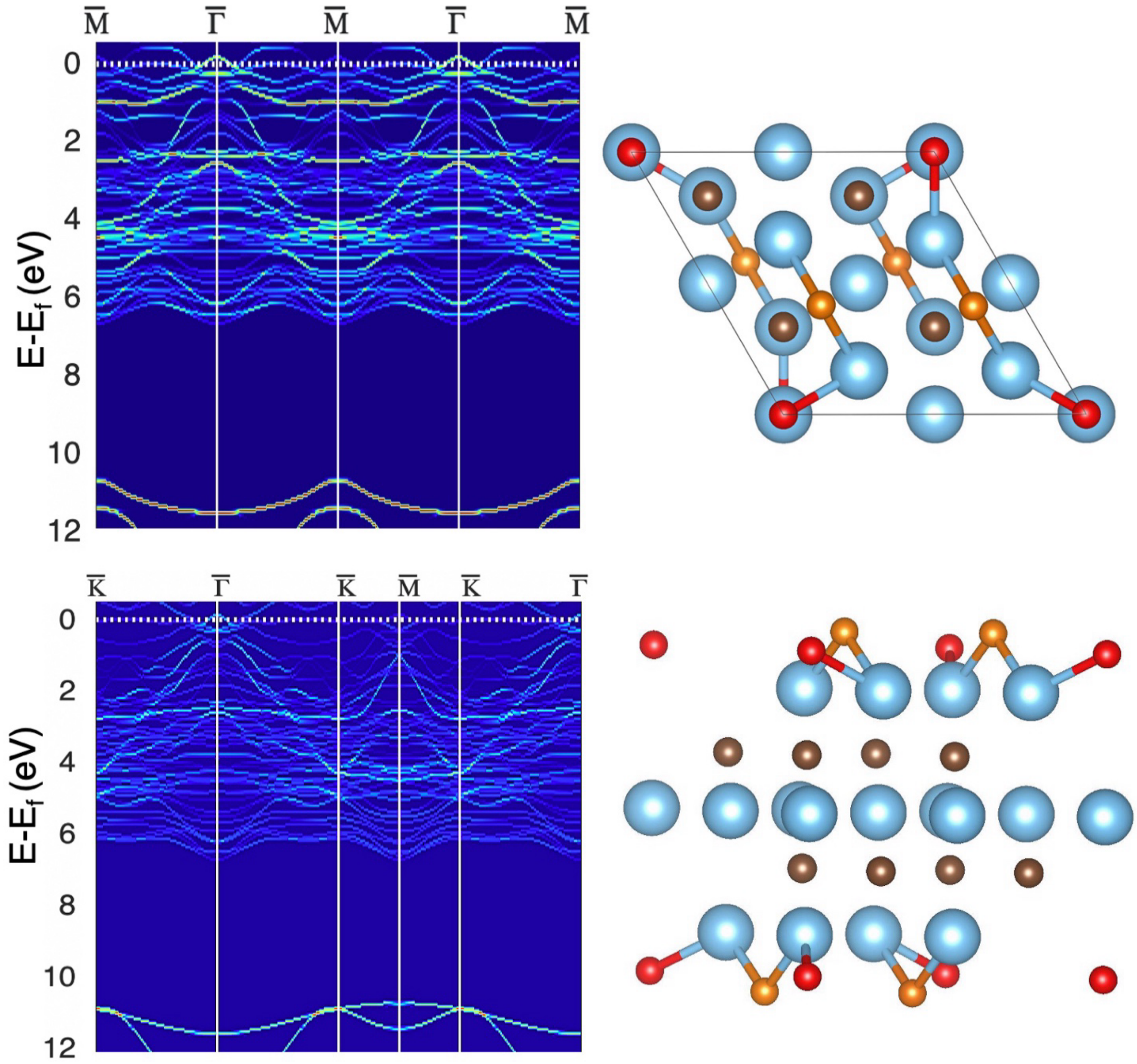

**Figure S10:** Unfolded bands along the $\bar{M}$-$\bar{\Gamma}$-$\bar{M}$-$\bar{\Gamma}$-$\bar{M}$ (top) and $\bar{K}$-$\bar{\Gamma}$-$\bar{K}$-$\bar{M}$-$\bar{K}$-$\bar{\Gamma}$ (bottom) directions of $Ti_3C_2O_AO_b$ with a 2x2 supercell model (right) shows a flat band at 1.5 eV along the $\bar{M}$-$\bar{\Gamma}$-$\bar{M}$ direction. The $O_{fcc}$ atoms are illustrated by red color, and the $O_b$ atoms by orange color.